\documentclass[fleqn,10pt]{wlscirep}
\usepackage[utf8]{inputenc}
\usepackage[T1]{fontenc}
\usepackage{multicol}
\usepackage{amssymb}
\title{Energy absorption in the laser-QED regime}

\author[1,*]{Alex F Savin}
\author[1]{Aimee J Ross}
\author[1]{Ramy Aboushelbaya}
\author[1]{Marko W Mayr}
\author[1]{Ben Spiers}
\author[1]{Robin H-W Wang}
\author[1,2]{Peter A Norreys}
\affil[1]{Clarendon Laboratory, University of Oxford, Parks Road, Oxford, OX1 3PU, United Kingdom}
\affil[2]{Central Laser Facility, STFC Rutherford Appleton Laboratory, Didcot, OX11 0QX, United Kingdom}

\affil[*]{alexander.savin@physics.ox.ac.uk}

\begin{abstract}
A theoretical and numerical investigation of non-ponderomotive absorption at laser intensities relevant to quantum electrodynamics is presented. It is predicted that there is a regime change in the dependence of fast electron energy on incident laser energy that coincides with the onset of pair production via the Breit-Wheeler process. This prediction is numerically verified via an extensive campaign of QED-inclusive particle-in-cell simulations. The dramatic nature of the power law shift leads to the conclusion that this process is a candidate for an unambiguous signature that future experiments on multi-petawatt laser facilities have truly entered the QED regime.
\end{abstract}
\begin{document}

\flushbottom
\maketitle
%
%
\begin{multicols}{2}
\section*{Introduction}
With the impending completion and commissioning of the Extreme Light Infrastructure - in particular ELI Beamlines\cite{MourouELI}, and the Apollon laser facility\cite{ZouApollon} amongst other multi-petawatt laser systems, it will soon be possible to investigate an entirely new regime of plasma physics in the laboratory. At this frontier of high-energy density physics it will be possible to conduct experimental investigations into a range of topics including non-linear quantum electrodynamic (QED) processes that spawn electron-positron pair production\cite{BellPairProduction} and laser wakefield acceleration of electron bunches up to multi-GeV energies\cite{MartinsLWFA}. There will also be the opportunity to make advances in several other fields within laser-plasma interactions such as coherent harmonic generation and focusing\cite{GordienkoFocusHHG,DromeyExp1,DromeyExp2}, attosecond science\cite{SadlerXFEL} ion beam characterisation and acceleration\cite{RobinsonRPA,HenigBOA,FiuzaCSA,ClarkIon,SnaveleyTNSA}, electron beam generation via laser-channeling and hole-boring\cite{CeurvorstChanneling,SarriHosing,SarriChannels}, and laboratory astrophysics\cite{MeszarosGRB}.

In order for future experiments to be able to claim with categorical certainty that results are due to QED effects, it is necessary that there be an easily verifiable signature that the laser-plasma interactions have entered the QED regime. While there has been significant work exploring electron-positron cascade production\cite{VranicCascades,GCasc}, and laser energy absorption\cite{GrismayerQEDAbs} far into the QED regime, in this paper, we consider the modifications to laser energy absorption on the cusp of entering the QED regime. In this paper, we suggest that such a signature could be found by considering the changes to absorption processes at extremely high laser intensities. It is already known that the processes by which plasma absorbs incident laser energy moves through several regimes as the incident laser intensity is increased\cite{WilksIntense}. These regimes are often identified by how the ``hot'' electron energy scales with the electron density, $n_e$, and/or the normalised amplitude of the laser, $a_0$:

\end{multicols}
\begin{figure}[b]
    \centering
    \includegraphics[scale=0.25]{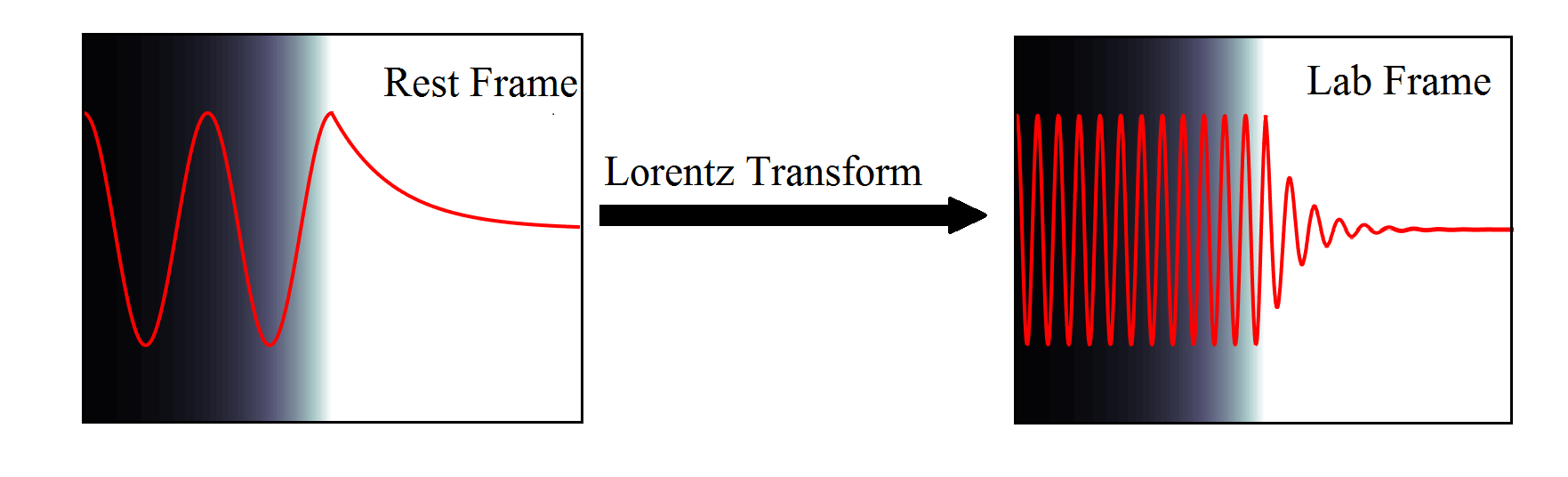}
    \caption{A simple schematic of how transforming from the rest frame of the expanding plasma into the laboratory frame leads to a continued propagation of the vector potential wave (red) beyond the plasma's critical surface - indicated by the boundary between dark (vacuum) and white (plasma).}
    \label{fig:RelTrans}
\end{figure}

\begin{figure}
    \centering
    \includegraphics[scale = 0.4]{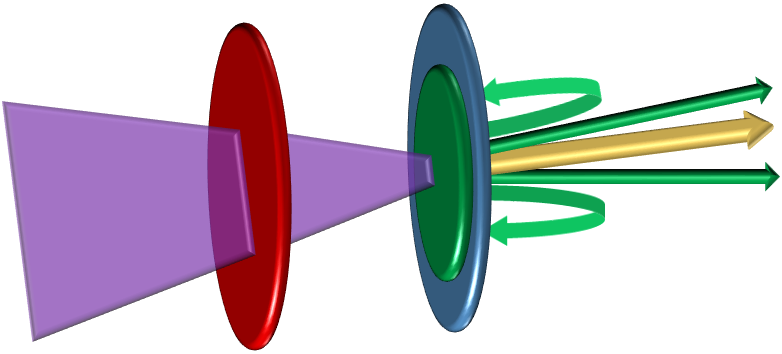}
    \caption{A pictorial representation of the proposed modification to the ZVP mechanism to include QED effects. The original pseudo-capacitor, characterised by a displaced electron fluid (blue) and a consequent region of net positive space charge (red) is augmented by pair production at the peak intensity of the incident laser pulse (faded purple) on the negative ``plate''. The positrons (yellow), not feeling a restoring force, propagate into the bulk plasma and leave the region of interest. In contrast, some of the pair-produced electrons (green), are subject to a restoring force which keeps a fraction of them within the region of interest, thus increasing the net charge of the negative ``plate''. }
    \label{fig:QEDpic}
\end{figure}

\begin{multicols}{2}
\begin{equation}
    a_0 = \frac{eA}{m_\mathrm{e}c} \equiv \sqrt{\frac{I_{\mathrm{W\,cm^{-2}}}\lambda^2_{\mathrm{\mu m}}}{1.37\times10^{18}}}
    \label{eq: a0}
\end{equation}
where $A$ is the vector potential, $I$ the intensity, and $\lambda$ the wavelength of the incident laser pulse, $c$ is the speed of light in vacuum, and $e$ and $m_\mathrm{e}$ are the electronic charge and rest mass.

At relatively modest laser intensities ($a_0<1$), the dominant absorption processes are inverse bremsstrahlung\cite{Colvin&Larsen} and resonant absorption\cite{ForslundAnharmonicResonance}. As the laser power is increased to relativistic intensities - identified by $a_0 > 1$ - a number of other absorption mechanisms have been proposed theoretically and verified experimentally. Notable, and dominant, examples include the explicitly ponderomotive mechanism\cite{WilksPond} and Brunel (or vacuum) heating\cite{BrunelHeating,GibbonVacuum}. Theoretical and numerical work in recent years has predicted that when moving into the regime of ``ultra-relativistic'' lasers incident on relativistically over-dense targets, absorption processes are dominated by non-ponderomotive mechanisms such as the zero-vector-potential (ZVP) absorption mechanism\cite{BaevaZVP,SavinZVP}. This regime is identified by $a_0 \geq 5$, and $n_\mathrm{e} \gg n_\mathrm{c}$ where $n_\mathrm{c}$ is the critical density of the plasma, above which the plasma is opaque to light of wavelengths less than or equal to $\lambda_0$:
\begin{equation}
    n_\mathrm{c} = \frac{4\pi^2m_\mathrm{e}\varepsilon_0c^2}{e^2}\frac{1}{\lambda^2_0} \equiv \frac{1.11\times10^{21}}{\lambda^2_{0\,\,\mathrm{\mu m}}}\,\mathrm{cm^{-3}}
    \label{eq: nc}
\end{equation}
Here, $\varepsilon_0$ is the permittivity of free space.

The non-relativistic, ponderomotive, and non-ponderomotive regimes are distinguished experimentally by the scaling of the ``hot electron'' energy with $a_0$. Specifically, $T\propto a_0^x$, where: $x < 1$ for the non-relativistic case\cite{ForslundAnharmonicResonance}, $x = 1$ in the ponderomotive regime\cite{WilksIntense}, and $x \approx 2$ in the non-ponderomotive regime\cite{BaevaZVP,SavinZVP}. In this paper, we propose an alteration to the ZVP model devised by Baeva \emph{et al.}\cite{BaevaZVP} that accounts for the possibility of generating electron-positron pairs and model how such pair production would affect the scaling of ``hot'' electron energy with $a_0$. By considering this energy scaling, it is suggested that laser-energy absorption by the plasma electrons offers a possible experimentally verifiable signature of entrance into the QED regime.

The structure of the paper is as follows: firstly, a theoretical justification for the ZVP's continued effect at QED-relevant intensities is considered. Next, a simple physical model capturing the dynamics of energy absorption is presented, including the effect of electron-positron pair production. Finally, the results of a series of particle-in-cell simulations are then presented which numerically validate the predictions of our model and pictorially demonstrate the signature we expect will indicate the entrance of future experimental campaigns into the QED regime.

\section*{Results}

\subsection*{ZVP at QED-relevant Intensities}

The existence of the ZVP mechanism relies on a shift in perception when changing between reference frames. If one considers a laser pulse incident on a plasma with an ablating front surface, there are two principal frames of interest. The first is the rest frame of the bulk plasma, or ``lab frame''; and the second is the rest frame of the expanding ablation front which will be referred to here as the ``rest frame''. In the rest frame, the incident laser radiation changes form at the critical density surface from a propagating oscillation to an evanescent decay\cite{BaevaZVP}. Previous work has established that for non-relativistic ablation velocities, for any angle of incidence, transforming into the lab frame yields a short skin depth, $\delta$, over which the oscillations in the laser pulse's vector potential propagate beyond the critical density surface\cite{SavinZVP}. As $\delta$ is independent of the decay length-scale of the ablation front, $\lambda_\mathrm{S}$, it is possible to enter the ZVP regime provided that $\lambda_\mathrm{S}$ is sufficiently small.

As laser intensities are increased, the velocity of the ablation front will also rise\cite{BataniAblation,FrantanduonoAblation}. As such, it is prudent to verify that the ZVP mechanism is still a viable absorption process at relativistic velocities. If one considers an obliquely incident laser pulse, its vector potential beyond the critical density surface in the rest frame, $\Vec{A}'_\mathrm{L}$ can be described as:
\begin{equation}
    \vec{A'_\mathrm{L}} = A_0'\cos{(\omega't')}\exp{(-r'/\lambda_\mathrm{S}')}\hat{r'}_\mathrm{pol}
    \label{eq: AL}
\end{equation}
where $A_0$ is the amplitude and $\omega$ the frequency of the laser pulse, $t$ is the time of propagation, $r'$ is the distance of propagation into the plasma, $\hat{r'}_\mathrm{pol}$ is the unit vector in the direction of polarisation of the laser pulse, and primed variables indicate that the quantity is defined and measured in the rest frame (consequentially, non-primed variables are defined and measured in the lab frame).

By constraining the problem to an s-polarised pulse propagating in the $x-y$ plane, with $\hat{x}$ being the axis of plasma ablation, the unit vector $\hat{r}_\mathrm{pol}$, in the lab frame can be written as:
\begin{equation}
    \hat{r}_\mathrm{pol} = \sin\theta\hat{x} - \cos\theta\hat{y}
    \label{eq: rpol}
\end{equation}
where $\theta$ is the angle of incidence of the laser pulse onto the expanding plasma. By considering the headlight effect\cite{SteaneHeadlight}, and the Lorentz transformation of the electromagnetic 4-potential\cite{Steane4vectors} (\emph{cf.} Methods Section), a series of deductions can be made about the vector potential in the lab frame:
\begin{eqnarray}
& A'\sin\theta'& =  A\sin\theta \\
\therefore & A'&  =  \gamma A(1- \beta\cos\theta)\\
\\
\implies & A & = \frac{A_0'\cos{(\omega't')}e^{-r'/\lambda_\mathrm{S}'}}{\gamma(1-\beta\cos\theta)}
\label{eq: Alab}
\end{eqnarray}
By applying a Lorentz transformation to the position 4-vector and converting variables to a pseudo-equivalent in the lab frame, the final form for $A$ can be deduced:
\begin{equation}
    A = A_0\cos{(\omega t - kx)}\exp{\left(-\frac{[(x-\beta ct)^2 +(y/\gamma)^2]^{1/2}}{\lambda_\mathrm{S}}\right)}
    \label{eq: Afinal}
\end{equation}
where $A_0 = A_0'/[\gamma(1-\beta\cos\theta)]$, $\omega = \gamma\omega'$, $k = \beta\gamma\omega/c$, $\lambda_\mathrm{S} = \lambda'_\mathrm{S}/\gamma$, and all coordinate variables are defined in their usual way as outlined in the Methods section.

It is clear from Equation \ref{eq: Afinal} that zeroes in the vector potential can continue to propagate into the over-dense plasma, even for relativistic ablation velocities, a simple schematic of this is shown in Figure \ref{fig:RelTrans}. With this principle established, it is possible now to consider how QED processes will impact the electron energy scaling with laser intensity.

\subsection*{New Energy Scaling}

In the original model for the ZVP absorption mechanism, the fast electron energy was deduced by assuming that the laser pressure displaces the electron fluid from the ionic background to set up a pseudo-capacitor system with the plates replaced by two charge fluids of charge $Q$, separated by a small displacement, $\Delta r$\cite{BaevaZVP,SavinZVP}. This model yielded a dependence of the fast electron energy on the laser amplitude of $T \propto a_0^2$, which was successfully numerically validated up to a value of $a_0 = 100$\cite{SavinZVP}.

Beyond $a_0 = 100$, QED effects -- such as pair production -- have the potential to contribute to laser-plasma interactions. As the intensity of the laser is further increased, the extent of pair production is predicted to increase logarithmically\cite{BellPairProduction}. Ultra-high intensity laser pulses are expected to act as a background electromagnetic field that can stimulate the Breit-Wheeler pair production process\cite{BreitWheeler}. When applied to the preexisting ZVP model, this leads to the assumption that at the point of the laser's peak intensity, the rate of pair production will be maximal. This position of peak intensity coincides with the position to which the electron fluid is displaced by the laser pulse, a representation of which can be seen in Figure \ref{fig:QEDpic}.

\end{multicols}
\begin{figure}[h!]
    \centering
    \includegraphics[scale = 0.35]{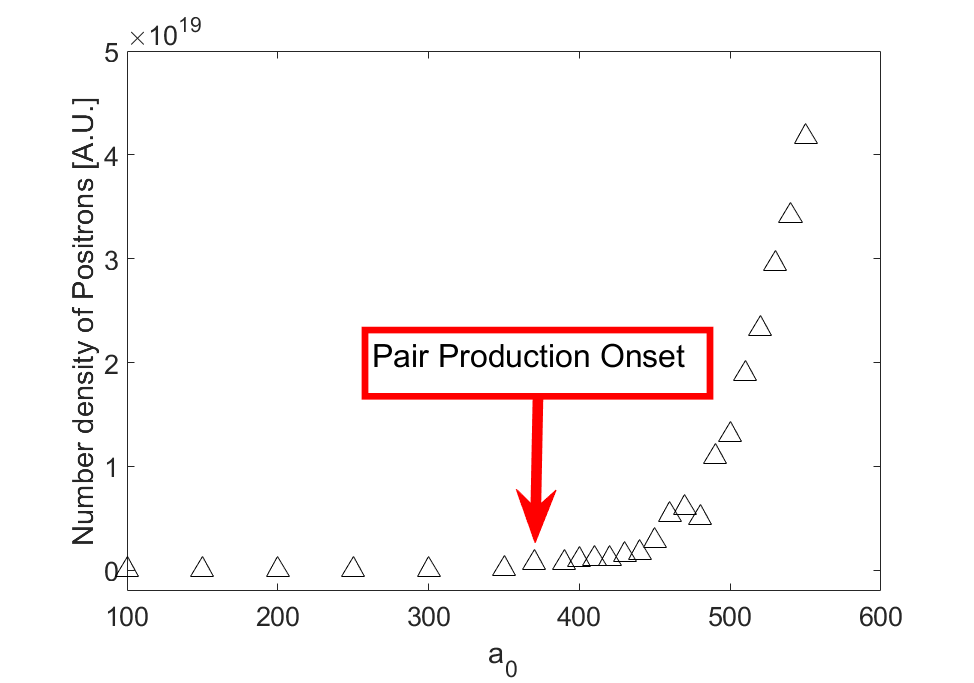}
    \caption{A plot of the number density of positrons scaling with the incident laser intensity, as extracted from particle-in-cell simulations. At $a_0 = 360$, pair production begins in earnest. The number density of positrons then increases rapidly with $a_0$.}
    \label{fig:posnum}
\end{figure}

\begin{figure}[h!]
    \centering
    \includegraphics[scale=0.5]{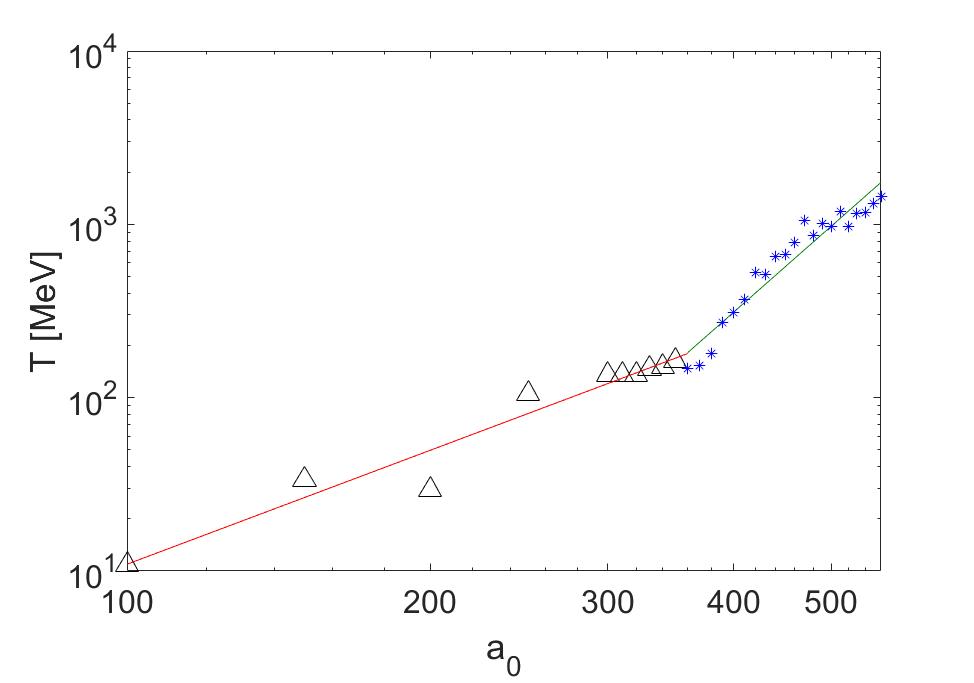}
    \caption{Plot of extracted fast electron energy against incident laser $a_0$. Black triangles indicate simulations for which no positrons were measured, the red line is a fit corresponding to $T\propto a_0^{2.18\pm0.18}$. The blue asterisks indicate results from simulations for which positrons were detected. The green line is a fit corresponding to $T\propto a_0^{5.12\pm1.34}$.}
    \label{fig:Tvsa}
\end{figure}
\begin{multicols}{2}

With electron-positron pairs being produced on one of the pseudo-capacitor's ``plates'', it is necessary to modify the ZVP model to account for these additional particles. As the e$^-$-\,e$^+$ pairs are produced at the negative plate, one would expect that the pairs will be subject to the electric field of the pseudo-capacitor. Thus, as the pairs will be created with high energy and momentum predominantly in the laser-forward direction, the positrons will propagate unhindered into the quasi-neutral bulk plasma.

In contrast, some number, $f$, of the newly created electrons could be captured onto the negative ``plate'' by the pseudo-capacitor's electric field, as shown in Figure \ref{fig:QEDpic}. This problem is best considered by comparing the kinetic energy, $T_\mathrm{esc}$, required to escape from the pseudo-capacitor's electric field to the average energy, $\langle U\rangle$, of the newly created electronic energy distribution. The escape energy will be of the same order of the energy that an electron would gain by crossing the capacitor, which was found -- in prior work\cite{BaevaZVP,SavinZVP} --  to be proportional to $a_0^2$, while numerical investigations into the Bethe-Heitler process\cite{BetheHeitler} have found that the average energy of pair-produced positrons (and by inference, pair-produced electrons) is proportional to $1/a_0$\cite{LobetEEEnergy}.

If one considers the pair-produced electronic energy distribution function $n(\epsilon)$, where $\epsilon$ is the ratio of energy to $\langle U\rangle$, then the number of captured electrons can be thought of as:
\begin{equation}
    f \propto n(\epsilon \leq T_\mathrm{esc}/\langle U\rangle)
    \label{eq: dtnfn}
\end{equation}
In the limit that the escape energy is much less than the average pair-produced electron energy, it is possible, in principle to Taylor expand the distribution function and by preserving only first-order terms, Equation \ref{eq: dtnfn} becomes:
\begin{equation}
    f \propto \frac{T_\mathrm{esc}}{\langle U\rangle} \equiv a_0^3
    \label{eq:fvsa}
\end{equation}

The final consideration, therefore, is to examine how the additional electrons, $f$, on the negative ``plate'' of the pseudo-capacitor impacts the energy scaling of the fast electrons. If the pseudo-capacitor has a cross-sectional area, $\sigma$, then the associated electric field in the non-QED case is given by $E = Q/\sigma\epsilon_0$. However, at QED-relevant intensities, the capacitor adds some number of electrons, $f$, to the negative ``plate'' such that while the charge of the positive ``plate'' remains as $Q$, the negative ``plate''' is modified to:
\begin{equation}
    Q'_\mathrm{neg} = -Q(1+f)
    \label{eq:negplate}
\end{equation}
Therefore, the pseudo-capacitor's electric field is modified to:
\begin{equation}
    E' = \frac{Q}{\sigma\epsilon_0}\left(1 + \frac{1}{2}f\right)
\end{equation}
Given that the kinetic energy, $T$, of the fast electrons is calculated by considering the energy gained as an electron crosses the pseudo-capacitor, $T$ can be expressed as:
\begin{equation}
    T = e\vec{E}\cdot\Delta\vec{r} \propto \frac{a_0^2}{n_e\lambda^2}\left(1+\frac{1}{2}f\right)
\end{equation}
By using the scaling relation derived in Equation \ref{eq:fvsa} this yields:
\begin{equation}
    T \propto \frac{a_0^5}{n_e}
    \label{eq: prediction}
\end{equation}
With a cubic increase in the dependence of $T$ on $a_0$ it is expected that on entering the QED regime, the change in electron energy scaling with incident laser intensity will become swiftly apparent.

\subsection*{Numerical Validation}

In lieu of experimental measurements to verify the above model, fully relativistic and kinetic particle-in-cell codes were used to run QED-inclusive simulations of intense laser pulses interacting with critically over-dense plasma.

The particle-in-cell codes OSIRIS\cite{FonsecaOSIRIS} and EPOCH\cite{ArberEPOCH} were both used to run a parameter scan of the fast electron energy as the value of $a_0$ was increased from 100 up to 560. The simulations were run in two dimensions, as the dynamics of the ZVP interaction are known to be confined to the plane of propagation of the laser pulse\cite{SavinZVP}. The simulations modelled an intense laser of varying $a_0$, with a wavelength of 1\,$\mathrm{\mu m}$, incident upon an aluminum plasma 1\,$\mathrm{\mu m}$ thick with an exponentially decaying density profile of decay-scale length 0.2\,$\mathrm{\mu m}$ on the laser-incident side. The density of the plasma was set to be $5.55\times10^{22}$\,cm$^{-3}$, corresponding to 50 times the critical density. A package comprising a QED Monte Carlo integrator\cite{RidgersEPOCHQED} was included in the simulation to model pair production. The energy of the fast electrons, and the number of positrons at the end of the simulation were extracted from each simulation run to determine the trend of fast electron energy and positron number against incident laser intensity.

As can be seen in Figure \ref{fig:posnum}, from $a_0 = 360$ upwards, there are a significant number of positrons detected in the simulation and the number rises rapidly with increasing laser intensity. Correspondingly, Figure \ref{fig:Tvsa} plots the fast electron energy against $a_0$. It is clear that around $a_0 = 350$, there is a distinct change in the dependence of $T$ on $a_0$. By dividing the acquired data into two sets: the first corresponding to when pair production is not observed, and the second corresponding to when pair production is prevalent, it is possible to obtain qualitative information about the power law linking $T$ and $a_0$. In the non-QED regime:
\begin{equation}
    T \propto a_0^{2.18\pm0.18}
    \label{eq: tvaclas}
\end{equation}
which is broadly in agreement with previous work investigating the ZVP mechanism\cite{BaevaZVP,SavinZVP}. In contrast, upon entering the QED regime, the dependence of the fast electron energy on the laser intensity changes dramatically to:
\begin{equation}
    T \propto a_0^{5.12\pm0.34}
    \label{eq: tvaqed}
\end{equation}
This is in excellent agreement with the prediction of Equation \ref{eq: prediction}.

\section*{Discussion}

We have predicted and numerically verified that there is a distinct regime change for the electronic absorption of incident laser energy that is coincident with the onset of quantum electrodynamics effects. The prediction that the power law linking fast electron energy and incident laser amplitude would change from quadratic to quintic was borne out by the results of an extensive simulation campaign. This leads us to the conclusion that QED effects such as pair production augment non-ponderomotive absorption mechanisms, such as the ZVP mechanism in a dramatic fashion.

Such a significant change in energy scaling, makes measurement of the ZVP mechanism a candidate for being a clear and unambiguous indicator that future high-power laser systems, such as the Extreme Light Infrastructure have truly breached the energy barrier into the QED regime. This clarity could allow experimental investigations into a wealth of unexplored physics to proceed with the confidence that quantum electrodynamics is truly playing a role in extreme high energy density physics.

\section*{Methods}

\subsection*{Relativistic dynamics}

The Lorentz transformation and headlight effect\cite{Steane4vectors,SteaneHeadlight} were used when transforming variables between the laboratory and rest frames detailed in the main text:
\begin{eqnarray*}
    A'^{\mu} & = & \Lambda^\mu_\nu A^\nu \\
    x'^{\mu} & = & \Lambda^\mu_\nu x^\nu \\
    \cos\theta' & = & \frac{\cos\theta - \beta}{1 - \beta\cos\theta} \\
    \sin\theta' & = & \frac{1}{\gamma}\frac{\sin\theta}{1 - \beta\cos\theta}
\end{eqnarray*}
where $\beta$ is the ratio of the ablation velocity to the speed of light in vacuum, $\gamma = [1 - \beta^2]^{-1/2}$ is the well-known Lorentz factor, $A^\mu = (\phi/c, A\cos\theta, A\sin\theta)$ is the electromagnetic 4-potential, $x^\mu = (ct, x, y)$ is the space-time 4-vector, and $\Lambda$ is the Lorentz transform given by:
 \begin{equation*}   
   \Lambda = \begin{pmatrix}
    \gamma & -\beta\gamma & 0 \\
    -\beta\gamma & \gamma & 0 \\
    0 & 0 & 1 
    \end{pmatrix}
\end{equation*}

\subsection*{Particle-in-cell simulations}

Both OSIRIS\cite{FonsecaOSIRIS} and EPOCH\cite{ArberEPOCH} operate according to the same principles of particle-in-cell (PIC) codes. Particles are modelled as ``macro-particles''. The negatively charged macro-particles can be considered to be electrons as they bear the same charge:mass ratio as electrons. The positively charged macro-particles were modelled to have the same charge:mass ratio as deuterons. PIC codes update the electromagnetic fields' spatial distributions and macro-particles' positions and trajectories on alternate time-steps, i.e.\,on every odd time-step, the fields are updated by solving Maxwell's equations, given the distribution of macro-particles. On each subsequent even time-step, the macro-particles are updated according to the Lorentz force law given the updated electromagnetic fields. QED packages include the addition of a Monte Carlo integrator\cite{RidgersEPOCHQED} to calculate the probabilities and rates of pair-production, amongst other QED effects.

One key difference between EPOCH and OSIRIS is that while EPOCH takes Syst\`eme International (SI) units, OSIRIS uses a dimensionless variation where all quantities are normalised to some user-defined reference time- and length-scales. For the benefits of discussion here, the quantities are listed in SI units for both codes, a simple numerical correction was computed for the OSIRIS parameters.

The total size of the simulation box was $20\,\mathrm{\mu m}\times20\,\mathrm{\mu m}$. 100 cells per micron were used to resolve the $\hat{x}$-direction (corresponding to the axis of laser propagation), and 25 cells per micron were used to resolve the $\hat{y}$-direction. 20 macro-ions and 500 macro-electrons were modelled within each cell. The laser was polarised in the $\hat{y}$-direction. The time-step increment of the simulations was set to 2.5\,as for a total duration of 200\,fs.

OSIRIS was used in prior work\cite{BaevaZVP,SavinZVP} and in this paper to obtain simulation data points from $100< a_0 < 250$. EPOCH was used to obtain data from $200<a_0<500$. The overlap between the OSIRIS simulations and prior work\cite{BaevaZVP,SavinZVP} was necessary to ensure the correct physics was being modelled by the simulations consistently. The overlap between the use of the OSIRIS and EPOCH codes was necessary to ensure that the two codes were consistently modelling the same situation without numerical errors.

\bibliography{ZVPQEDbib}

\begin{thebibliography}{10}
\urlstyle{rm}
\expandafter\ifx\csname url\endcsname\relax
  \def\url#1{\texttt{#1}}\fi
\expandafter\ifx\csname urlprefix\endcsname\relax\def\urlprefix{URL }\fi
\expandafter\ifx\csname doiprefix\endcsname\relax\def\doiprefix{DOI: }\fi
\providecommand{\bibinfo}[2]{#2}
\providecommand{\eprint}[2][]{\url{#2}}

\bibitem{MourouELI}
\bibinfo{author}{Mourou, G.~A.}, \bibinfo{author}{Labaune, C.~L.},
  \bibinfo{author}{Dunne, M.}, \bibinfo{author}{Naumova, N.} \&
  \bibinfo{author}{Tikhonchuk, V.~T.}
\newblock \bibinfo{journal}{\bibinfo{title}{Relativistic laser-matter
  interaction: from attosecond pulse generation to fast ignition}}.
\newblock {\emph{\JournalTitle{Plasma Phys. Contr. F.}}}
  \textbf{\bibinfo{volume}{49}}, \bibinfo{pages}{B667} (\bibinfo{year}{2007}).

\bibitem{ZouApollon}
\bibinfo{author}{Zou, J.} \emph{et~al.}
\newblock \bibinfo{journal}{\bibinfo{title}{Design and current progress of the
  {A}pollon {10 PW} project}}.
\newblock {\emph{\JournalTitle{High Power Lasr Science and Engineering}}}
  \textbf{\bibinfo{volume}{3}}, \bibinfo{pages}{e2} (\bibinfo{year}{2015}).

\bibitem{BellPairProduction}
\bibinfo{author}{Bell, A.} \& \bibinfo{author}{Kirk, J.~G.}
\newblock \bibinfo{journal}{\bibinfo{title}{Possibility of {P}rolific {P}air
  {P}roduction with {H}igh-{P}ower {L}asers}}.
\newblock {\emph{\JournalTitle{Phys. Rev. Lett.}}}
  \textbf{\bibinfo{volume}{101}}, \bibinfo{pages}{200403}
  (\bibinfo{year}{2008}).

\bibitem{MartinsLWFA}
\bibinfo{author}{Martins, S.~F.} \emph{et~al.}
\newblock \bibinfo{journal}{\bibinfo{title}{Numerical simulations of {LWFA} for
  the next generation of laser systems}}.
\newblock {\emph{\JournalTitle{Advanced Accelerator Concepts}}}
  \textbf{\bibinfo{volume}{1086}}, \bibinfo{pages}{285} (\bibinfo{year}{2009}).

\bibitem{GordienkoFocusHHG}
\bibinfo{author}{Gordienko, S.}, \bibinfo{author}{Pukhov, A.},
  \bibinfo{author}{Shorokhov, O.}, \bibinfo{author}{Baeva, T.} \&
  \bibinfo{author}{Gordienko, T.}
\newblock \bibinfo{journal}{\bibinfo{title}{Coherent focusing of high
  harmonics: {A} new way towards the extreme intensities}}.
\newblock {\emph{\JournalTitle{Phys. Rev. Lett.}}}
  \textbf{\bibinfo{volume}{94}}, \bibinfo{pages}{103903}
  (\bibinfo{year}{2005}).

\bibitem{DromeyExp1}
\bibinfo{author}{Dromey, B.} \emph{et~al.}
\newblock \bibinfo{journal}{\bibinfo{title}{High harmonic generation in the
  relativistic limit}}.
\newblock {\emph{\JournalTitle{Nat. Phys.}}} \textbf{\bibinfo{volume}{2}},
  \bibinfo{pages}{456} (\bibinfo{year}{2006}).

\bibitem{DromeyExp2}
\bibinfo{author}{Dromey, B.} \emph{et~al.}
\newblock \bibinfo{journal}{\bibinfo{title}{Bright multi-ke{V} harmonic
  generation from relativistically oscillating plasma surfaces}}.
\newblock {\emph{\JournalTitle{Phys. Rev. Lett}}}
  \textbf{\bibinfo{volume}{99}}, \bibinfo{pages}{115003}
  (\bibinfo{year}{2007}).

\bibitem{SadlerXFEL}
\bibinfo{author}{Sadler, J.} \emph{et~al.}
\newblock \bibinfo{journal}{\bibinfo{title}{Compression of {X}-ray free
  electron laser pulses to attosecond duration}}.
\newblock {\emph{\JournalTitle{Sci. Rep.}}} \textbf{\bibinfo{volume}{5}},
  \bibinfo{pages}{16755} (\bibinfo{year}{2015}).

\bibitem{RobinsonRPA}
\bibinfo{author}{Robinson, A. P.~L.}, \bibinfo{author}{Zepf, M.},
  \bibinfo{author}{Kar, S.}, \bibinfo{author}{Evans, R.~G.} \&
  \bibinfo{author}{Bellei, C.}
\newblock \bibinfo{journal}{\bibinfo{title}{Radiation pressure acceleration of
  thing foils with circularly polarized laser pulses}}.
\newblock {\emph{\JournalTitle{New J. Phys.}}} \textbf{\bibinfo{volume}{10}},
  \bibinfo{pages}{013021} (\bibinfo{year}{2008}).

\bibitem{HenigBOA}
\bibinfo{author}{Henig, A.} \emph{et~al.}
\newblock \bibinfo{journal}{\bibinfo{title}{Enhanced laser-{D}riven
  {A}cceleration in the {R}elativstic {T}ransparency {R}egime}}.
\newblock {\emph{\JournalTitle{Phys. Rev. Lett.}}}
  \textbf{\bibinfo{volume}{103}}, \bibinfo{pages}{045002}
  (\bibinfo{year}{2009}).

\bibitem{FiuzaCSA}
\bibinfo{author}{Fiuza, F.} \emph{et~al.}
\newblock \bibinfo{journal}{\bibinfo{title}{Laser-{D}riven {S}hocl
  {A}cceleration of {M}onoenergetic {I}on {B}eams}}.
\newblock {\emph{\JournalTitle{Phys. Rev. Lett.}}}
  \textbf{\bibinfo{volume}{109}}, \bibinfo{pages}{215001}
  (\bibinfo{year}{2012}).

\bibitem{ClarkIon}
\bibinfo{author}{Clark, E.~L.} \emph{et~al.}
\newblock \bibinfo{journal}{\bibinfo{title}{Measurements of energetic proton
  transport through magnetized plasma from intense laser interactions with
  solids}}.
\newblock {\emph{\JournalTitle{Phys. Rev. Lett.}}}
  \textbf{\bibinfo{volume}{84}}, \bibinfo{pages}{670} (\bibinfo{year}{2000}).

\bibitem{SnaveleyTNSA}
\bibinfo{author}{Snavely, R.~A.} \emph{et~al.}
\newblock \bibinfo{journal}{\bibinfo{title}{Intense {H}igh-{E}nergy {P}roton
  {B}eams from {P}etawatt-{L}aser {I}rradiation of {S}olids}}.
\newblock {\emph{\JournalTitle{Phys. Rev. Lett.}}}
  \textbf{\bibinfo{volume}{85}}, \bibinfo{pages}{2945--2948}
  (\bibinfo{year}{2000}).

\bibitem{CeurvorstChanneling}
\bibinfo{author}{Ceurvorst, L.} \emph{et~al.}
\newblock \bibinfo{journal}{\bibinfo{title}{Channel optimization og
  high-intensity laser beams in millimeter-scale plasmas}}.
\newblock {\emph{\JournalTitle{Phys. Rev. E}}} \textbf{\bibinfo{volume}{97}},
  \bibinfo{pages}{043208} (\bibinfo{year}{2018}).

\bibitem{SarriHosing}
\bibinfo{author}{Sarri, G.} \emph{et~al.}
\newblock \bibinfo{journal}{\bibinfo{title}{Creation of persistent, straight, 2
  mm long laser driven channels in underdense plasmas}}.
\newblock {\emph{\JournalTitle{Phys. Plasmas}}} \textbf{\bibinfo{volume}{17}},
  \bibinfo{pages}{113303} (\bibinfo{year}{2010}).

\bibitem{SarriChannels}
\bibinfo{author}{Sarri, G.} \emph{et~al.}
\newblock \bibinfo{journal}{\bibinfo{title}{Observation of postsoliton
  expansion following laser propagation through an underdense plasma}}.
\newblock {\emph{\JournalTitle{Phys. Rev. Lett.}}}
  \textbf{\bibinfo{volume}{105}}, \bibinfo{pages}{175007}
  (\bibinfo{year}{2010}).

\bibitem{MeszarosGRB}
\bibinfo{author}{M\'esz\'aros, P.}
\newblock \bibinfo{journal}{\bibinfo{title}{Theories of gamma-ray bursts}}.
\newblock {\emph{\JournalTitle{Annu. Rev. Astron. Astrophys.}}}
  \textbf{\bibinfo{volume}{40}}, \bibinfo{pages}{137--169}
  (\bibinfo{year}{2002}).

\bibitem{VranicCascades}
\bibinfo{author}{Vranic, M.}, \bibinfo{author}{Grismayer, T.},
  \bibinfo{author}{Fonseca, R.~A.} \& \bibinfo{author}{Silva, L.~O.}
\newblock \bibinfo{journal}{\bibinfo{title}{Electron-positron cascades in
  multiple-laser optical traps}}.
\newblock {\emph{\JournalTitle{Plasma Phys. Control. Fus.}}}
  \textbf{\bibinfo{volume}{59}}, \bibinfo{pages}{014040}
  (\bibinfo{year}{2017}).

\bibitem{GCasc}
\bibinfo{author}{Grismayer, T.}, \bibinfo{author}{Vranic, M.},
  \bibinfo{author}{Martins, J.~L.}, \bibinfo{author}{Fonseca, R.~A.} \&
  \bibinfo{author}{Silva, L.~O.}
\newblock \bibinfo{journal}{\bibinfo{title}{Seeded qed cascades in
  counterpropagating laser pulses}}.
\newblock {\emph{\JournalTitle{Phys. Rev. E}}} \textbf{\bibinfo{volume}{95}},
  \bibinfo{pages}{023120} (\bibinfo{year}{2017}).

\bibitem{GrismayerQEDAbs}
\bibinfo{author}{Grismayer, T.}, \bibinfo{author}{Vranic, M.},
  \bibinfo{author}{Martins, J.~L.}, \bibinfo{author}{Fonseca, R.~A.} \&
  \bibinfo{author}{Silva, L.~O.}
\newblock \bibinfo{journal}{\bibinfo{title}{Laser absorption via quantum
  electrodynamics cascades in counter propagating laser pulses}}.
\newblock {\emph{\JournalTitle{Phys. Plasmas}}} \textbf{\bibinfo{volume}{23}},
  \bibinfo{pages}{056706} (\bibinfo{year}{2016}).

\bibitem{WilksIntense}
\bibinfo{author}{Wilks, S.~C.}, \bibinfo{author}{Kruer, W.~L.},
  \bibinfo{author}{Tabak, M.} \& \bibinfo{author}{Langdon, A.~B.}
\newblock \bibinfo{journal}{\bibinfo{title}{Absorption of ultra-intense laser
  pulses}}.
\newblock {\emph{\JournalTitle{Phys. Rev. Lett.}}}
  \textbf{\bibinfo{volume}{69}}, \bibinfo{pages}{1383} (\bibinfo{year}{1992}).

\bibitem{Colvin&Larsen}
\bibinfo{author}{Colvin, J.} \& \bibinfo{author}{Larsen, J.}
\newblock \emph{\bibinfo{title}{Extreme Physics: Properties and Behaviour of
  Matter at Extreme Conditions}} (\bibinfo{publisher}{Cambridge University
  Press}, \bibinfo{year}{2014}).

\bibitem{ForslundAnharmonicResonance}
\bibinfo{author}{Forslund, W.}, \bibinfo{author}{Kindel, J.~M.} \&
  \bibinfo{author}{Lee, K.}
\newblock \bibinfo{journal}{\bibinfo{title}{Theory of {H}ot {E}lectron
  {S}pectra at {H}igh {L}aser {I}ntensity}}.
\newblock {\emph{\JournalTitle{Phys. Rev. Lett.}}}
  \textbf{\bibinfo{volume}{39}}, \bibinfo{pages}{284} (\bibinfo{year}{1977}).

\bibitem{WilksPond}
\bibinfo{author}{Wilks, S.~C.} \& \bibinfo{author}{Kruer, W.~L.}
\newblock \bibinfo{journal}{\bibinfo{title}{Absorption of {U}trashort,
  {U}ltra-intense {L}aser {L}ight by {S}olids and {O}verdense {P}lasmas}}.
\newblock {\emph{\JournalTitle{IEEE J Quantum Elect.}}}
  \textbf{\bibinfo{volume}{33}}, \bibinfo{pages}{1954} (\bibinfo{year}{1997}).

\bibitem{BrunelHeating}
\bibinfo{author}{Brunel, F.}
\newblock \bibinfo{journal}{\bibinfo{title}{Not-{S}o-{R}esonant, {R}esonant
  {A}bsorption}}.
\newblock {\emph{\JournalTitle{Phys. Rev. Lett.}}}
  \textbf{\bibinfo{volume}{59}}, \bibinfo{pages}{52} (\bibinfo{year}{1987}).

\bibitem{GibbonVacuum}
\bibinfo{author}{Gibbon, P.} \& \bibinfo{author}{Bell, A.~R.}
\newblock \bibinfo{journal}{\bibinfo{title}{Collisionless {A}bsorption in
  {S}harp-{E}dged {P}lasmas}}.
\newblock {\emph{\JournalTitle{Phys. Rev. Lett.}}}
  \textbf{\bibinfo{volume}{68}}, \bibinfo{pages}{1535} (\bibinfo{year}{1992}).

\bibitem{BaevaZVP}
\bibinfo{author}{Baeva, T.}, \bibinfo{author}{Gordienko, S.},
  \bibinfo{author}{Robinson, A.} \& \bibinfo{author}{Norreys, P.}
\newblock \bibinfo{journal}{\bibinfo{title}{The zero vector potential mechanism
  of attosecond absorption}}.
\newblock {\emph{\JournalTitle{Phys. Plasma}}} \textbf{\bibinfo{volume}{18}},
  \bibinfo{pages}{056702} (\bibinfo{year}{2011}).

\bibitem{SavinZVP}
\bibinfo{author}{Savin, A.~F.} \emph{et~al.}
\newblock \bibinfo{journal}{\bibinfo{title}{Attosecond-scale absorption at
  extreme intensities}}.
\newblock {\emph{\JournalTitle{Phys. Plasma}}} \textbf{\bibinfo{volume}{24}},
  \bibinfo{pages}{113103} (\bibinfo{year}{2017}).

\bibitem{BataniAblation}
\bibinfo{author}{Batani, D.}
\newblock \bibinfo{journal}{\bibinfo{title}{Short-pulse laser ablation of
  materials at high intensities: Influence of plasma effects}}.
\newblock {\emph{\JournalTitle{Laser Part. Beams}}}
  \textbf{\bibinfo{volume}{28}}, \bibinfo{pages}{235--244}
  (\bibinfo{year}{2010}).

\bibitem{FrantanduonoAblation}
\bibinfo{author}{Frantanduono, D.~E.} \emph{et~al.}
\newblock \bibinfo{journal}{\bibinfo{title}{The direct measurement of ablation
  pressure driven by 351-nm laser radiation}}.
\newblock {\emph{\JournalTitle{J. Appl. Phys.}}}
  \textbf{\bibinfo{volume}{110}}, \bibinfo{pages}{073110}
  (\bibinfo{year}{2011}).

\bibitem{SteaneHeadlight}
\bibinfo{author}{Steane, A.~M.}
\newblock \emph{\bibinfo{title}{Relativity made relatively easy}}
  (\bibinfo{publisher}{OUP, United Kingdom}, \bibinfo{year}{2012}).

\bibitem{Steane4vectors}
\bibinfo{author}{Steane, A.~M.}
\newblock \emph{\bibinfo{title}{Relativity made relatively easy}}
  (\bibinfo{publisher}{OUP, United Kingdom}, \bibinfo{year}{2012}).

\bibitem{BreitWheeler}
\bibinfo{author}{Breit, G.} \& \bibinfo{author}{Wheeler, J.~A.}
\newblock \bibinfo{journal}{\bibinfo{title}{Collision of two light quanta}}.
\newblock {\emph{\JournalTitle{Phys. Rev.}}} \textbf{\bibinfo{volume}{46}},
  \bibinfo{pages}{1087--1091} (\bibinfo{year}{1934}).

\bibitem{BetheHeitler}
\bibinfo{author}{Bethe, H.} \& \bibinfo{author}{Heitler, W.}
\newblock \bibinfo{journal}{\bibinfo{title}{On the stopping of fast particles
  and on the creation of positive electrons}}.
\newblock {\emph{\JournalTitle{Proc. R. Soc. A}}}
  \textbf{\bibinfo{volume}{146}}, \bibinfo{pages}{83--112}
  (\bibinfo{year}{1934}).

\bibitem{LobetEEEnergy}
\bibinfo{author}{Lobet, M.}, \bibinfo{author}{Davoine, X.},
  \bibinfo{author}{d'Humi\`eres, D.} \& \bibinfo{author}{Gremillet, L.}
\newblock \bibinfo{journal}{\bibinfo{title}{Generation of high-energy
  electron-positron pairs in the collision of a laser-accelerated electron beam
  with a multipetawatt laser}}.
\newblock {\emph{\JournalTitle{Phys. Rev. Accel. Beams}}}
  \textbf{\bibinfo{volume}{20}}, \bibinfo{pages}{043401}
  (\bibinfo{year}{2017}).

\bibitem{FonsecaOSIRIS}
\bibinfo{author}{Fonseca, R.~A.} \emph{et~al.}
\newblock \bibinfo{journal}{\bibinfo{title}{{OSIRIS}: {A} three-dimensional,
  fully relativistic particle in cell code for modelling plasma based
  accelerators}}.
\newblock {\emph{\JournalTitle{Lect. Notes. Comp. Sci.}}}
  \textbf{\bibinfo{volume}{2231}}, \bibinfo{pages}{342} (\bibinfo{year}{2002}).

\bibitem{ArberEPOCH}
\bibinfo{author}{Arber, T.~D.} \emph{et~al.}
\newblock \bibinfo{journal}{\bibinfo{title}{Contemporary particle-in-cell
  approach to laser-plasma modelling}}.
\newblock {\emph{\JournalTitle{Plasma Phys. Control. Fus.}}}
  \textbf{\bibinfo{volume}{57}}, \bibinfo{pages}{113001}
  (\bibinfo{year}{2015}).

\bibitem{RidgersEPOCHQED}
\bibinfo{author}{Ridgers, C.~P.} \emph{et~al.}
\newblock \bibinfo{journal}{\bibinfo{title}{Modelling gamma-ray photon emission
  and pair production in high-intensity laser-matter interactions}}.
\newblock {\emph{\JournalTitle{J. Comp. Phys.}}}
  \textbf{\bibinfo{volume}{260}}, \bibinfo{pages}{273--285}
  (\bibinfo{year}{2014}).

\end{thebibliography}

\section*{Acknowledgements}

A.F.S and P.A.N. thank Professor Chris Ridgers for useful discussions. They also thank Professor Lu\`is O. Silva, Thomas Grismayer, and Marija Vranic for invaluable support with the OSIRIS code. Funding for this work was provided by the Science and Technology Facilities Council of the United Kingdom and EPSRC under Grant Nos. ST/P000967/1 and EP/N509711/1. A.F.S. acknowledges support from RCUK under student number 1796896. The authors are grateful for computing resources provided by STFC Computing Department's SCARF cluster. This work also used the ARHER UK National Computing Service under EPSRC grant number EP/R029148/1. The authors thank the OSIRIS Consortium for access to the OSIRIS PIC code. Access to EPOCH was provided thanks to the UK EPSRC funded Collaborative Computational Project in Plasma Physics - grant reference number EP/M022463/1.

\section*{Author Contributions}

A.F.S. undertook the mathematical study of both the viability of the ZVP mechanism at QED-relevant intensities and the modification to the energy scaling relation, and wrote the manuscript. A.F.S. and A.J. R. conducted the simulations that validate the theoretical model. R.A., M.W.M., and R.H-W.W., contributed to vital discussions necessary to develop the model and its validation. P.A.N. supervised the research.

\section*{Competing Interests}

The authors declare no competing interests.

\section*{Data Availability}

The OSIRIS particle-in-cell code is available on application to the Osiris consortium at: \hyperlink{www.picksc.idre.ucla.edu}{www.picksc.idre.ucla.edu}. The EPOCH particle-in-cell code is available on application via the Collaborative Computational Project in Plasma Physics at: \hyperlink{http://www.ccpp.ac.uk/codes.html}{http://www.ccpp.ac.uk/codes.html}. The input files for all simulations and output data files are available on request to A.F.S. E-mail: alexander.savin@physics.ox.ac.uk.

\end{multicols}
\end{document}